\documentclass[prl,superscriptaddress,twocolumn,showpacs]{revtex4}

\usepackage{graphicx}
\usepackage{times}
\usepackage{amsmath,amssymb}

\newcommand{\field}{$\langle 111 \rangle$\ }

\begin{document}

\title{Possible proximity of the Mott insulating Iridate Na$\bf _2$IrO$\bf _3$ to a topological phase:\\
        Phase diagram of the Heisenberg-Kitaev model in a magnetic field}

\author{Hong-Chen Jiang}
\affiliation{Microsoft Research, Station Q, University of California, Santa Barbara, CA 93106}

\author{Zheng-Cheng Gu}
\affiliation{Kavli Institute for Theoretical Physics, University of California, Santa Barbara, CA 93106}

\author{Xiao-Liang Qi}
\affiliation{Microsoft Research, Station Q, University of California, Santa Barbara, CA 93106}
\affiliation{Department of Physics, Stanford University, Stanford, CA 94305 }

\author{Simon Trebst}
\affiliation{Microsoft Research, Station Q, University of California, Santa Barbara, CA 93106}

\date{\today}

\begin{abstract}
Motivated by the recent experimental observation of a Mott insulating state for the layered Iridate
Na$_2$IrO$_3$, we discuss possible ordering states of the effective Iridium moments in
the presence of strong spin-orbit coupling and a magnetic field.
For a field pointing in the \field direction
-- perpendicular to the hexagonal lattice formed by the Iridium moments --
we find that a combination of Heisenberg and Kitaev exchange interactions gives
rise to a rich phase diagram with both symmetry breaking magnetically ordered phases
as well as a topologically ordered phase that is stable over a small range of coupling parameters.
Our numerical simulations further indicate two exotic multicritical points at the boundaries
between these ordered phases.
\end{abstract}


\maketitle


In the realm of condensed matter physics, spin-orbit coupling has long been considered a residual, relativistic correction
of minor relevance to the macroscopic properties of a material. In recent years this perspective has dramatically changed,
especially due to the theoretical prediction and subsequent experimental observation of fundamentally new states of
quantum matter, so-called topological insulators~\cite{TopologicalInsulators}, that are solely due to the effect of spin-orbit coupling.
The topological insulators experimentally realized so far are semiconductors, whose physical properties can be largely captured
by band theory of non-interacting electrons.
It is an interesting challenge, for both theory and experiment, to identify an even broader class of materials where
this physics plays out even in the presence of interactions and strong correlations \cite{TopologicalMottInsulators}.
Good candidate materials for the latter are the Iridates \cite{Iridates,Gegenwart}.
These $5d$ transition metal oxides are prone to exhibit
electronic correlations and form (weak) Mott insulators, while the relatively large mass of the Iridium ions ($Z=77$) gives rise
to a comparably strong spin-orbit coupling, which has been found to be as large as $\lambda \approx 400$~meV \cite{IridiumSOEstimate}.
The most common valence of the Iridium ions in these materials is Ir$^{4+}$. The $d$-orbitals of this $5d^5$ configuration are
typically split by the surrounding crystal field, and for the octahedral geometry of the IrO$_6$ oxygen cage, result in an orbital configuration
where five electrons occupy the lowered, threefold degenerate $t_{\rm 2g}$ level.
Spin-orbit coupling will further lift this degeneracy of the $t_{\rm 2g}$ orbitals and for strong coupling the effective $l=1$ orbital angular
momentum~\cite{t2g-Orbitals} is combined with the $s=1/2$ spin degree of freedom carried by the hole of this partially filled $t_{\rm 2g}$ orbital
configuration. This leaves us with two Kramers doublets of total angular momentum $j=3/2$ and $j=1/2$, of which the former is
of lower energy and fully occupied by four electrons, while the partial filling of the latter gives rise to an effective spin-1/2 degree of freedom.

\begin{figure}[t]
   \centerline{
    \includegraphics[width=\columnwidth]{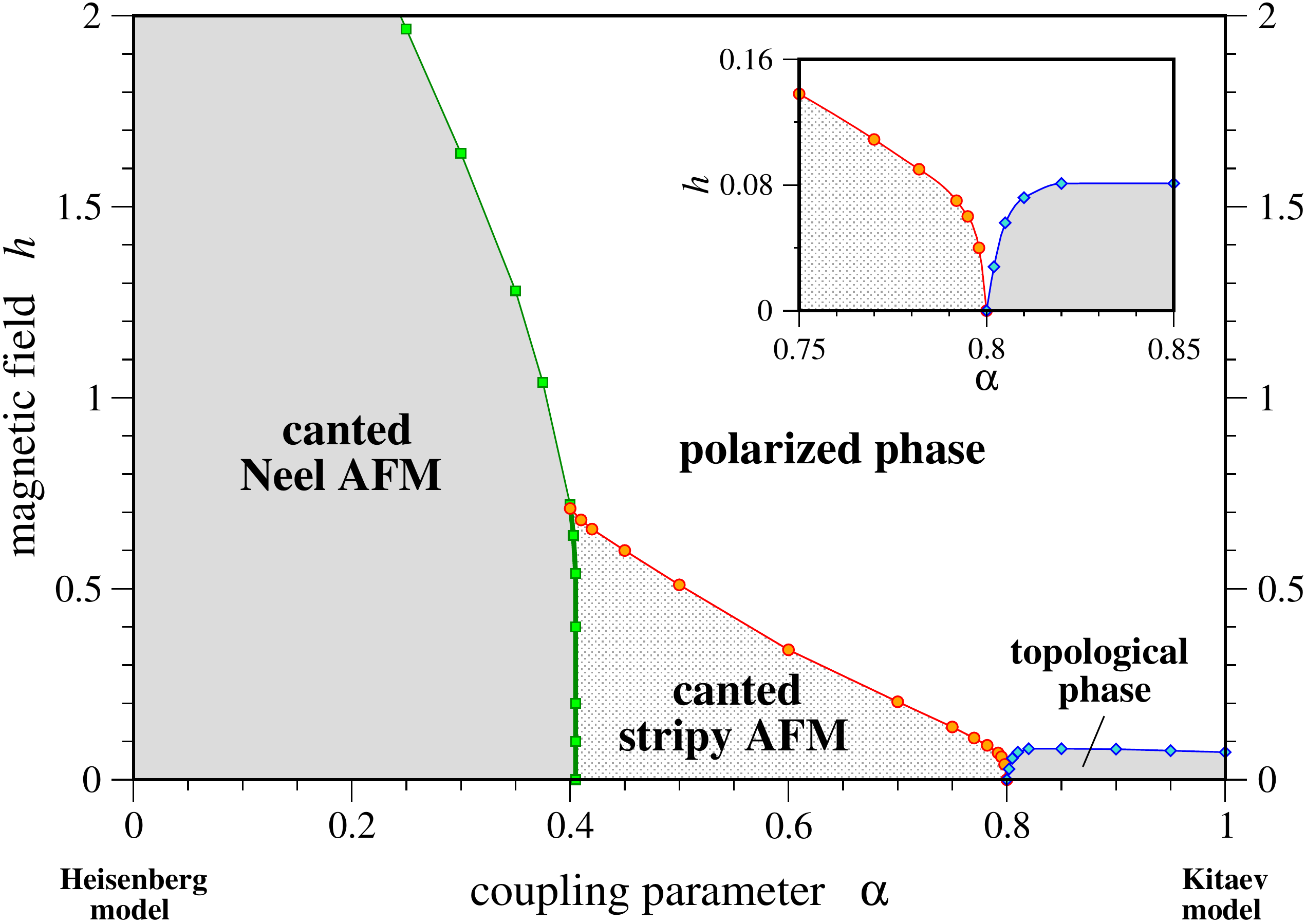}
    }
   \caption{(color online)
              Ground-state phase diagram of the Heisenberg-Kitaev model
              (\ref{Eq:KitaevHeisenbergHamiltonian}) in a \field magnetic
              field of strength $h$.
              Interpolating from the Heisenberg ($\alpha=0$) to Kitaev ($\alpha=1$)
              limit for small field strength, a sequence of three ordered phases is observed:
              a canted N\'eel state for $\alpha \lesssim 0.4$, a canted stripy N\'eel state
              illustrated in Fig.~\ref{Fig:LatticeOrder}c) for $0.4 \lesssim \alpha \lesssim 0.8$,
              and a topologically ordered state for non-vanishing field around the Kitaev limit.
              All ordered phases are destroyed for sufficiently large magnetic
              field giving way to a polarized state.
             }
   \label{Fig:Phase_HK}
\end{figure}

In this manuscript we focus on the Iridate Na$_2$IrO$_3$, in which NaIr$_2$O$_6$ slabs are stacked along the crystallographic $c$-axis,
and the Ir$^{4+}$ ions in the layers form a hexagonal lattice  \cite{Gegenwart}.
Recent measurements of the magnetic susceptibility provide evidence of effective spin-1/2 moments and magnetic
correlations below $T_N \approx 15$~K indicating that Na$_2$IrO$_3$ is indeed a Mott insulator  \cite{Gegenwart}.
Theoretically, it has been argued  \cite{Jackeli09,Chaloupka10} that the interactions between the effective Iridium moments in the
Mott regime are captured by a combination of isotropic and highly anisotropic exchanges, which can be tracked back to the spin and
orbital components of the effective momenta. A microscopic Hamiltonian interpolating between these two types of exchanges
is given by
\begin{equation}
   H_{\rm HK} =  (1-\alpha) {\sum_{\langle i,j\rangle}} \vec{\sigma}_i\cdot \vec{\sigma}_j
                        - 2\alpha \sum_{{\gamma \rm-links}} {\sigma_i^{\gamma} \sigma_j^{\gamma}} \,,
   \label{Eq:KitaevHeisenbergHamiltonian}
\end{equation}
where the $\sigma_i$ denote the effective spin-1/2 moment of the Ir$^{4+}$ ions, $\gamma = x,y,z$ indicates
the three different links of the hexagonal lattice, and $0 \leq \alpha \leq 1$ parametrizes the relative coupling
strength of the isotropic and anisotropic exchange between the moments.
For $\alpha=0$ the Hamiltonian reduces to the ordinary Heisenberg model, while in the opposite limit of
highly anisotropic exchanges ($\alpha=1$) the system corresponds to the Kitaev model \cite{Kitaev2006}.
The latter is known to exhibit a gapless spin-liquid ground state (for equal coupling along the links)
that can be gapped out into a topological phase with non-Abelian quasiparticle excitations
by certain time-reversal symmetry breaking perturbations
\cite{Kitaev2006}.
One such perturbation is a magnetic field pointing in the \field direction, perpendicular to the honeycomb layer
\begin{eqnarray}
  H_{\rm HK+h} = H_{\rm HK} - {\sum_i}\vec{h}\cdot \vec{\sigma_i} \,.
  \label{Eq:Hamiltonian_field}
\end{eqnarray}
The main result of our manuscript is the rich phase diagram of this model, shown in Fig.~\ref{Fig:Phase_HK}.
Besides two conventional, magnetically ordered phases we find a topologically ordered phase and two multicritical
points, which we will discuss in detail in the remainder of the manuscript.

\begin{figure}
   \centerline{
    \includegraphics[width=\columnwidth]{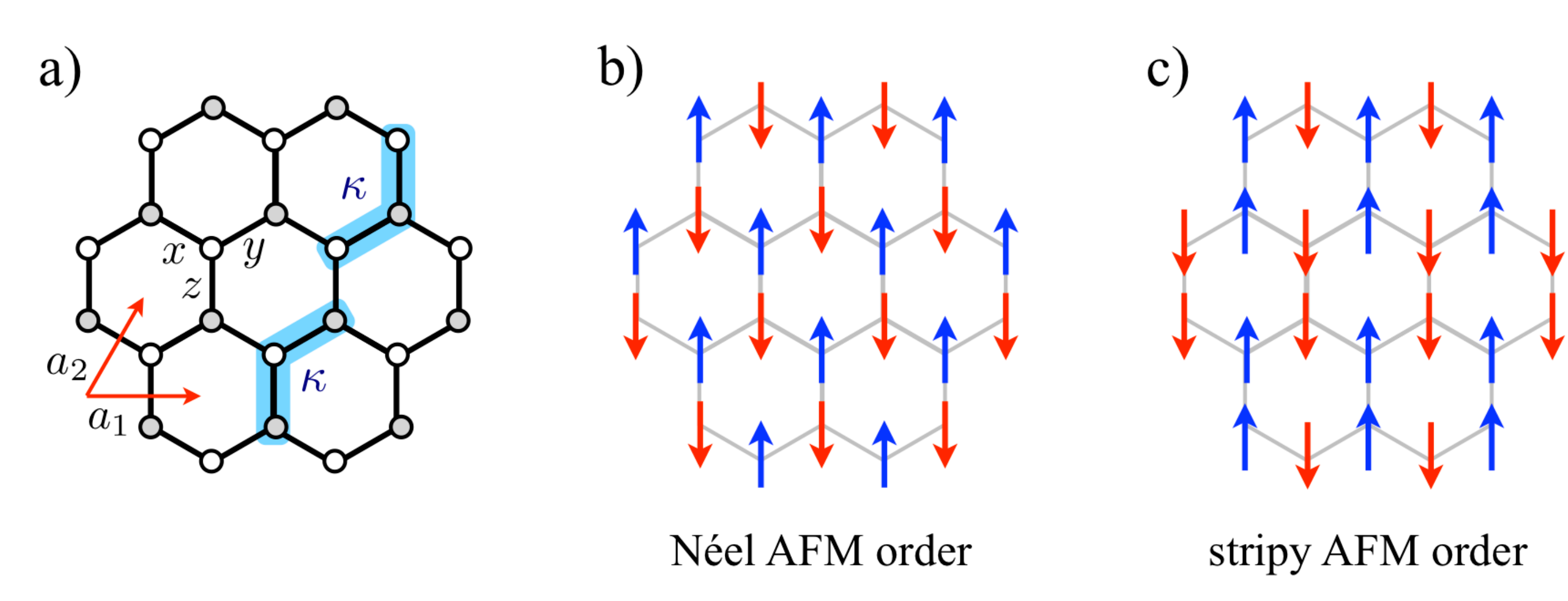}
    }
   \caption{(color online)
                 a) The honeycomb lattice spanned by unit vectors
                      $\vec{a}_1 = (1,0)$ and $\vec{a}_2 = (1/2, \sqrt{3}/2)$.
                 Illustration of magnetic states with b) Neel order and
                 c) stripy Neel order.}
   \label{Fig:LatticeOrder}
\end{figure}

\paragraph{Numerical simulations.--}
We determine the ground-state phase diagram of Hamiltonian \eqref{Eq:Hamiltonian_field}
by extensive `quasi-2D' density-matrix renormalization group (DMRG) \cite{White1992} calculations
on systems with up to $N=64$ sites. In particular, we consider clusters of size $N=2\times N_1\times N_2$,
which are spanned by multiples $N_1 \vec{a}_1$ and $N_2 \vec{a}_2$ of the unit cell
vectors $\vec{a}_1 = (1,0)$ and $\vec{a}_2 = (1/2, \sqrt{3}/2)$ as illustrated in Fig.~\ref{Fig:LatticeOrder}.
It should be noted that the numerical analysis of Hamiltonian \eqref{Eq:Hamiltonian_field} is a challenging endeavor,
since not only the entire Hilbert space needs to be considered (due to the lack of SU(2) invariance),
but one also has to work with complex data types (due to the \field orientation of the magnetic field).
Our DMRG calculations keep up to $m=2048$ states,
which is found to give excellent convergence with typical truncation errors of
less than $10^{-8}$.
We further use periodic boundary conditions in both lattice directions, which reduces
finite-size effects.
We have determined the phase boundaries in Fig.~\ref{Fig:Phase_HK} by extensive scans
of the ground-state energy, magnetization, and their derivatives in the ($\alpha, h$)-parameter
space~\cite{FootNoteTRG}.


\paragraph*{Magnetically ordered states.--}%
We start our discussion of the phase diagram shown in Fig.~\ref{Fig:Phase_HK} by first recapitulating previous results
\cite{Chaloupka10} for the Heisenberg-Kitaev model \eqref{Eq:KitaevHeisenbergHamiltonian}
in the absence of a magnetic field.
Interpolating the relative coupling strength $\alpha$ between the isotropic Heisenberg limit ($\alpha=0$) and
the highly anisotropic Kitaev limit ($\alpha=1$) a sequence of three phases has been observed \cite{Chaloupka10}:
The N\'eel ordered state of the Heisenberg limit is stable for $\alpha \lesssim 0.4$, when it gives way to a `stripy'
N\'eel ordered state illustrated in Fig.~\ref{Fig:LatticeOrder} which covers the coupling regime
$0.4 \lesssim \alpha \lesssim 0.8$.
In the extended parameter regime $0.8 \lesssim \alpha \leq 1$ the collective ground state is
a gapless spin liquid. Near $\alpha=1$, perturbation theory reveals that the gapless excitations of this phase are
emergent 
Majorana fermions forming two Dirac cones in momentum space.

Including a magnetic field in the \field direction a rich phase diagram evolves out of
this sequence of three phases. For the magnetically ordered states we find that the orientation
of the order in the N\'eel and stripy AFM phase cants along the \field direction.
To further characterize these canted states, it is helpful to analyze the independent symmetries
of Hamiltonian \eqref{Eq:Hamiltonian_field}.
Besides the lattice translational symmetry $T$ and a reflection symmetry $I$ around the
centers of the hexagons,
there is an additional $C_3^*$ symmetry, which is a combination of a three-fold rotation
around an arbitrary lattice site and a three-fold spin rotation along the \field spin axis
\cite{FootNoteCyclicSymmetry}.
Both canted phases break a subset of these discrete symmetries of the Hamiltonian.
The canted N\'eel order breaks the $C_3^*$ and the $I$ symmetries, which thus leads to
a six-fold ground-state degeneracy in this phase.
The canted stripy phase breaks both the $C_3^*$ and translational
symmetry (since the ordering pattern doubles the unit cell).
As a consequence, we also find a six-fold ground-state degeneracy
in this phase.

\begin{figure}[b]
\centerline{
    \includegraphics[width=\columnwidth]{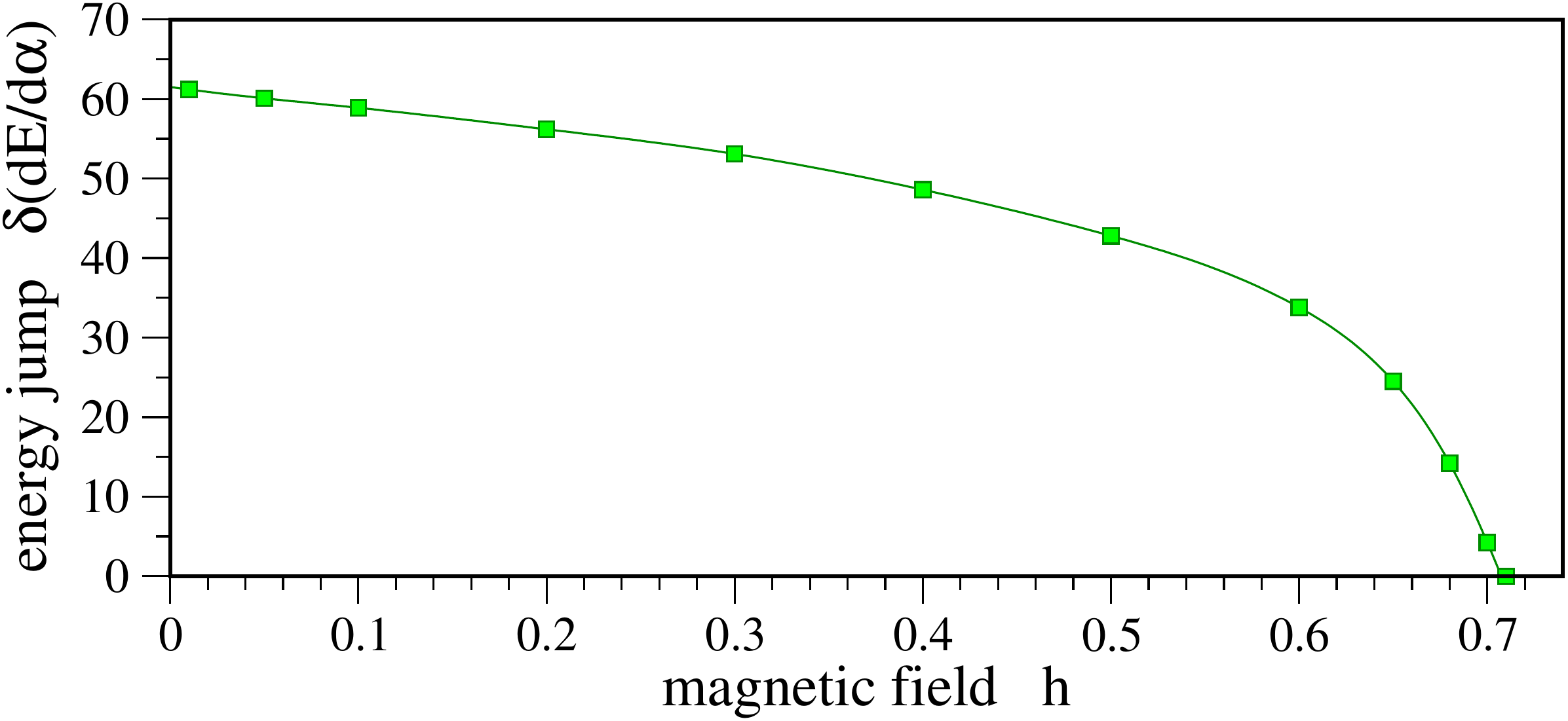}
    }
\caption{(color online) Energy jump along the first-order transition
between the canted N\'eel and stripy AFM.}\label{Fig:jump}
\end{figure}

For sufficiently large magnetic field, the order of both canted phases is destroyed
and they give way to a simple polarized state. Our numerical simulations strongly suggest
that the transitions between the polarized state and these canted states are continuous,
which is in agreement with their spontaneous symmetry breaking.
On the other hand, the transition between the two canted states at finite field strength
(indicated by the bold line in Fig.~\ref{Fig:Phase_HK}) is found to be first-order.
In our simulations this is indicated by a sharp drop of the first derivative of the
energy $dE/d\alpha$ as a function of the coupling parameter $\alpha$ across this
transition -- as shown in Fig.~\ref{Fig:jump} for increasing strength of the magnetic
field $h$. Approaching the endpoint of this first-order line around $h_c\simeq0.7$ this drop
smoothly vanishes, which indicates that this endpoint possibly is a tricritical point
at which the two canted magnetically ordered phases and the polarized phase meet.
The existence of such a tricritical point can be understood within a Landau description
with two distinct order parameters -- corresponding to the discrete symmetry breaking
of the two different magnetically ordered phases -- opening a gap to magnon excitations.

\begin{figure}[t]
  \begin{center}
    \includegraphics[width=\columnwidth]{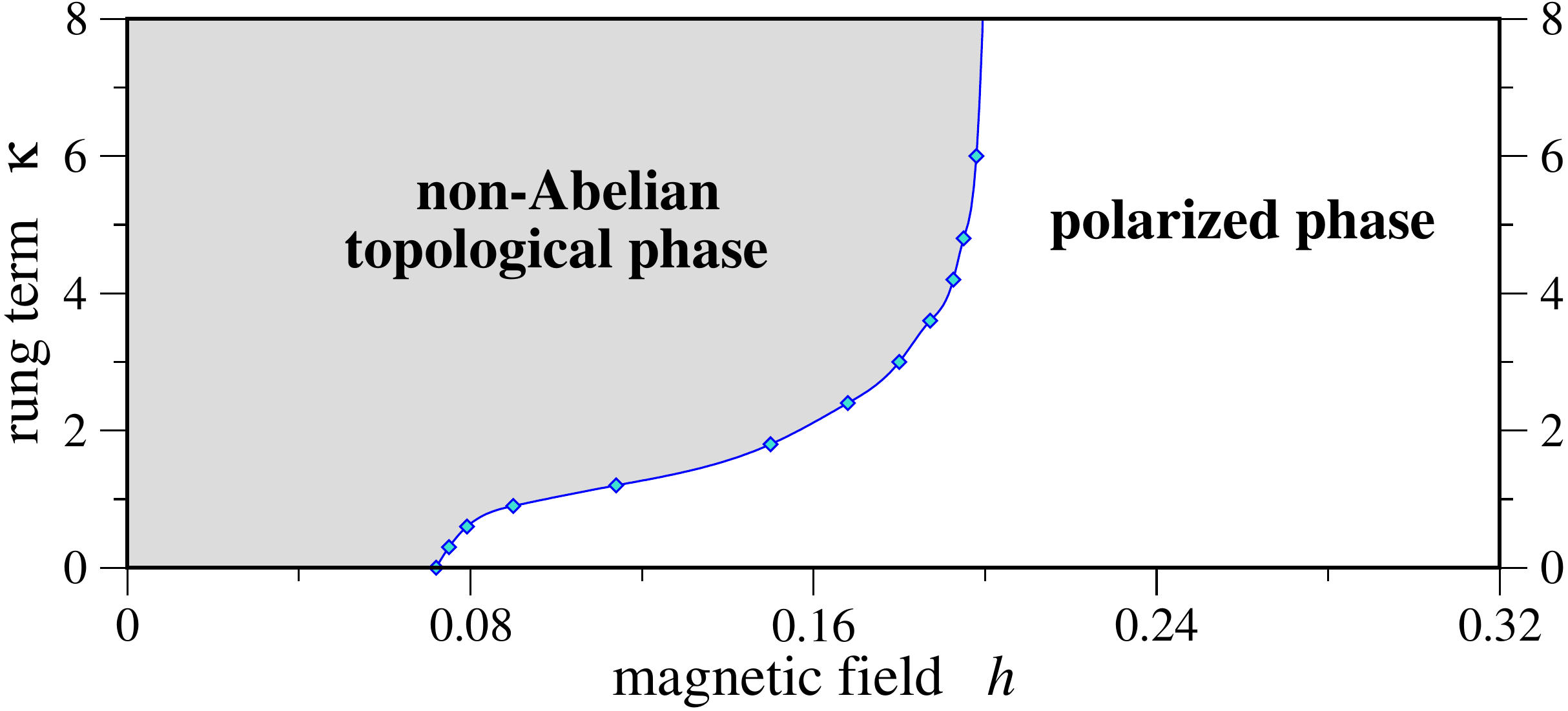}
    \vskip 2mm
    \includegraphics[width=\columnwidth]{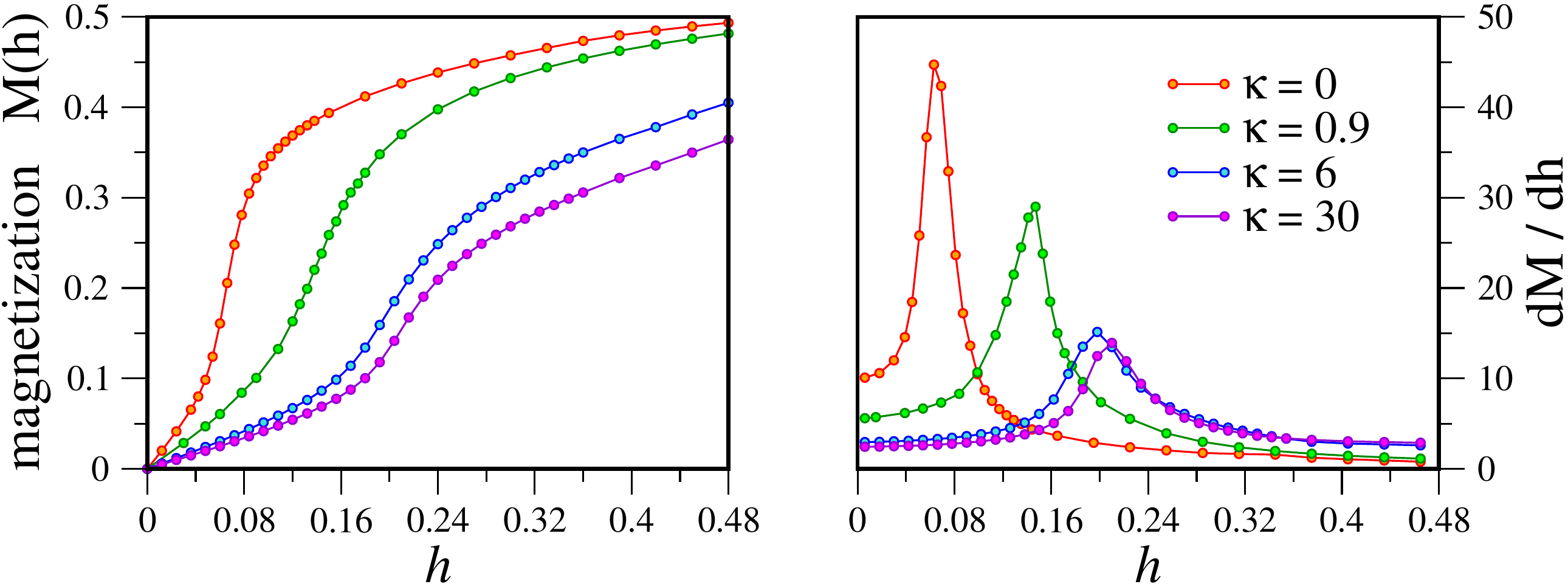}
  \end{center}
  \caption{(color online)
              Top panel: Ground-state phase diagram of the Kitaev model ($\alpha=1$)
              in the $h-\kappa$ plane, where $h$ is the strength of a magnetic
              field pointing in the \field direction and $\kappa$ is the strength of
              a time-reversal symmetry breaking three-site term.
              Lower panel: Magnetization sweeps and its derivative for various
              $\kappa$.}
  \label{Fig:Phase_KRung}
\end{figure}

\paragraph*{Topological phase.--}
We now turn to the spin liquid phase found for coupling parameters
$0.8 \lesssim \alpha \leq 1$. For the Kitaev limit ($\alpha=1$) it has previously
been argued \cite{Kitaev2006} that an infinitesimal field along the \field direction
will drive the system from the gapless spin liquid into a gapped non-Abelian topologically
ordered phase. As we will discuss in the following our numerical simulations
allow us to confirm the existence of such a topologically ordered state for small
magnetic field strengths not only in the Kitaev limit, but for the full extent of the
gapless spin liquid phase, as indicated in the phase diagram of Fig.~\ref{Fig:Phase_HK}.
We will further present an independent and non-perturbative way to determine
the topological nature of this phase. This complements the original argument
by Kitaev  \cite{Kitaev2006}, which was primarily based on a perturbation expansion
showing that the leading order effect of a small magnetic field $h$ is to introduce a
topological mass term for the Majorana fermions -- however, such a perturbative
argument should be carefully tested when applied to a gapless state.
For this purpose, we consider an additional three-spin exchange term $\kappa$,
indicated by the blue bonds in Fig. \ref{Fig:LatticeOrder}a), in our Hamiltonian
\begin{equation}
    H_{\rm HK+h+\kappa} = H_{\rm HK+h} - \kappa{\sum_{ijk}}\sigma^x_i \sigma^y_j \sigma^z_k \,.
    \label{Eq:Hamiltonian_KRung}
\end{equation}
In the Kitaev limit ($\alpha=1$, $h=0$) this Hamiltonian is exactly solvable
in the same Majorana fermion representation used in the solution of the
unperturbed Kitaev model \cite{Kitaev2006}. In particular, one can prove
that the three-spin exchange $\kappa$ breaks time-reversal symmetry
and gaps out the spin liquid phase into a topologically ordered state with
non-Abelian excitations, so-called Ising anyons.
To demonstrate that a small magnetic field in the \field direction drives
the system into the same phase, we have numerically calculated the phase
diagram in the presence of both perturbations as shown in Fig.~\ref{Fig:Phase_KRung}.
The phase boundaries were again obtained by scanning the derivatives of ground-state
energy and magnetization in the $(h,\kappa)$-parameter space.
In particular, this phase diagram shows that one can adiabatically connect the phase
for large $\kappa$ and vanishing magnetic field with the phase for small,
non-vanishing magnetic field and $\kappa=0$, thus proving that the magnetic
field gaps out the spin liquid into the same non-Abelian topological phase
stabilized by the three-spin exchange.
The only feature in the diagram is a single phase transition line which separates the
topologically ordered state from the fully polarized state expected
for large magnetic field strengths. For the Kitaev limit ($\kappa=0$) this
transition occurs for $h_c\simeq 0.072$.
It is interesting to note that the critical field $h_c$ initially grows with increasing $\kappa$,
but then saturates to some finite value around $\kappa \gtrsim 6$. Physically, this saturation
can be understood by the behavior of the gap for the Majorana fermions in
the exact solution for $h=0$. The dispersion of the Majorana fermion is given by
$E_{\bf k}=2\sqrt{\left|1+e^{i{\vec{k} \cdot \vec{a}_1}}+e^{i{\vec{k} \cdot \vec{a}_2}}\right|^2+\kappa^2\sin^2({\vec{k} \cdot \vec{a}_1})}$.
For small $\kappa \ll 1$, the Majorana fermion has a gap $E_{\rm g}\simeq\sqrt{3}\kappa$.
However, for large $\kappa\gg 1$ the gap of Majorana fermion remains finite and independent from $\kappa$,
given by $E_{\rm g}\simeq 2$.
Since the magnetic field strength $h_c$ required to destroy the topological phase is determined by the
Majorana fermion gap at $h=0$, the critical field $h_c$ thus also increases and then saturates at large $\kappa$.

\begin{figure}[t]
  \centerline{
    \includegraphics[width=\columnwidth]{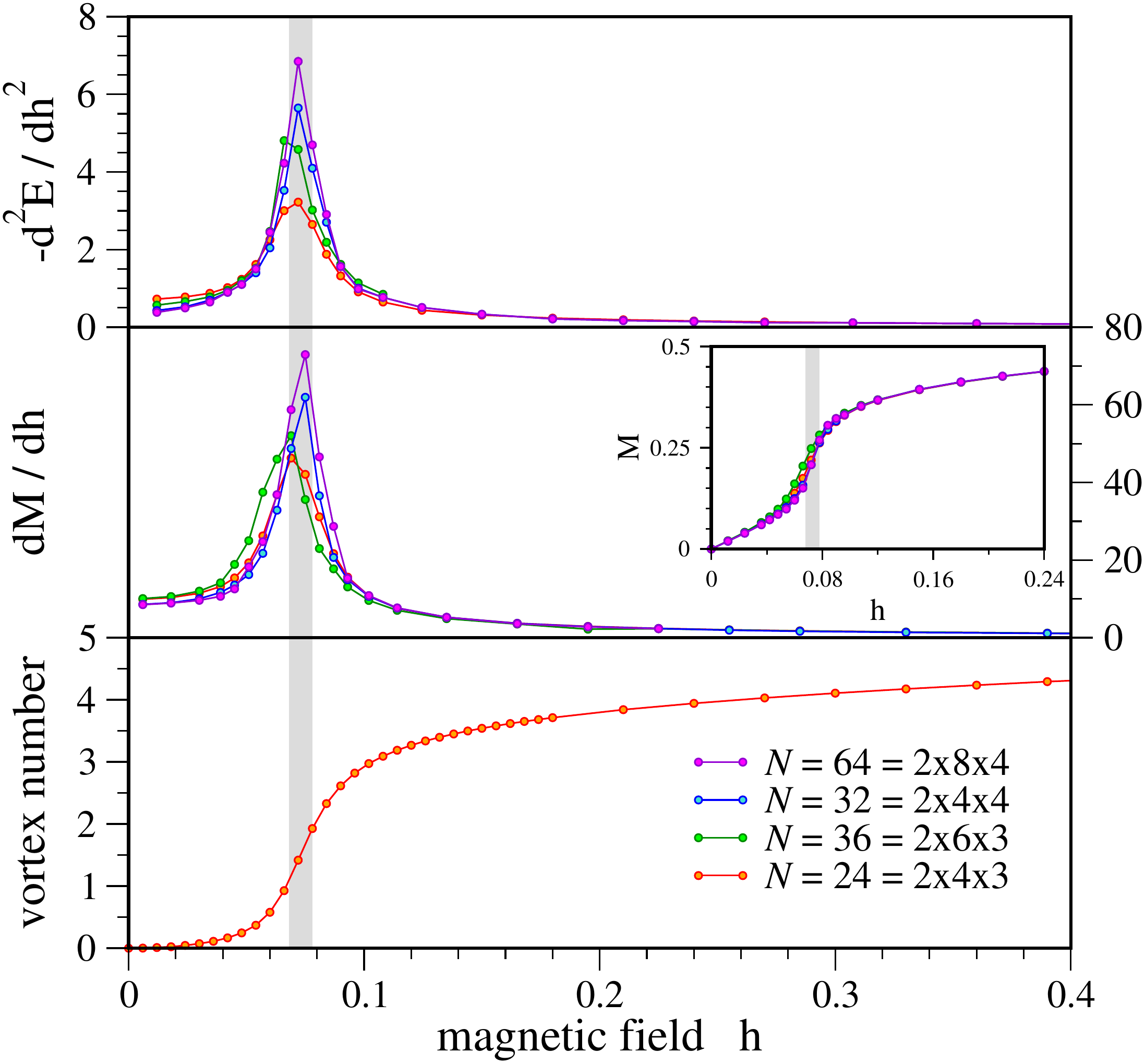}
    }
  \caption{(color online)
  	        Phase transition for the Kitaev model ($\alpha=1$) in a \field
	        magnetic field of strength $h$.
	        a) Second derivative of the ground state energy $-d^2E/dh^2$,
	        b) Magnetization $M(h)$ and its first derivative $dM(h)/dh$
	            for different system sizes.
	         c) Vortex number.}
  \label{Fig:dE2_Mh111}
\end{figure}

\paragraph*{Field-driven transition out of the topological phase.--}
We now return to the phase diagram of the Heisenberg-Kitaev model
in Fig.~\ref{Fig:Phase_HK} and focus on the transition between the
topologically ordered state and the polarized state for large field strength.
For the Kitaev limit ($\alpha=1$) this transition occurs at a critical
field strength of $h_c \approx 0.072$ and remains almost constant
as the coupling parameter $\alpha$ is decreased.
Interestingly, our numerics suggest that this field-driven phase transition
might be continuous or weakly first-order. In particular, we find that the
second-derivative of the ground state energy $-d^2E/dh^2$ at this transition
diverges with increasing system size, while the magnetization $M(h)$ does
not show any discontinuity, as shown in Fig.~\ref{Fig:dE2_Mh111}a) and b),
respectively.

While the limited system sizes in our study do not allow to unambiguously
determine the continuous nature of this field-driven phase transition, our numerics
nevertheless provide some further insights what might cause such a continuous
transition \cite{Gils09}. To this end, we plot the number of vortices in the ground state
as a function of magnetic field,  i.e. the number of plaquettes with a non-trivial flux,
in Fig.~\ref{Fig:dE2_Mh111}c).
Below the critical magnetic field, i.e. $h<h_c$, there are no vortices
indicating a deconfined phase
as expected in the presence of a vortex gap.
At the phase transition, however, the vortices appear to condense and the number
of vortices in the ground state quickly increase above the critical field strength.
The nature of the phase transition might thus be framed in terms of
a confinement-deconfinement transition of a non-Abelian gauge field, akin
to the confinement-deconfinement transition in the Abelian discrete gauge theory
\cite{FootNoteFieldStrength}.
An example of the latter is the $Z_2$ gauge theory, e.g. the toric code in a magnetic
field \cite{ToricCodeTransitions}, for which it is well known that flux condensation leads to a confinement
transition \cite{FradkinShenker}.


\paragraph*{A second multicritical point.--}

\begin{figure}[t]
   \centerline{
      \includegraphics[width=\columnwidth]{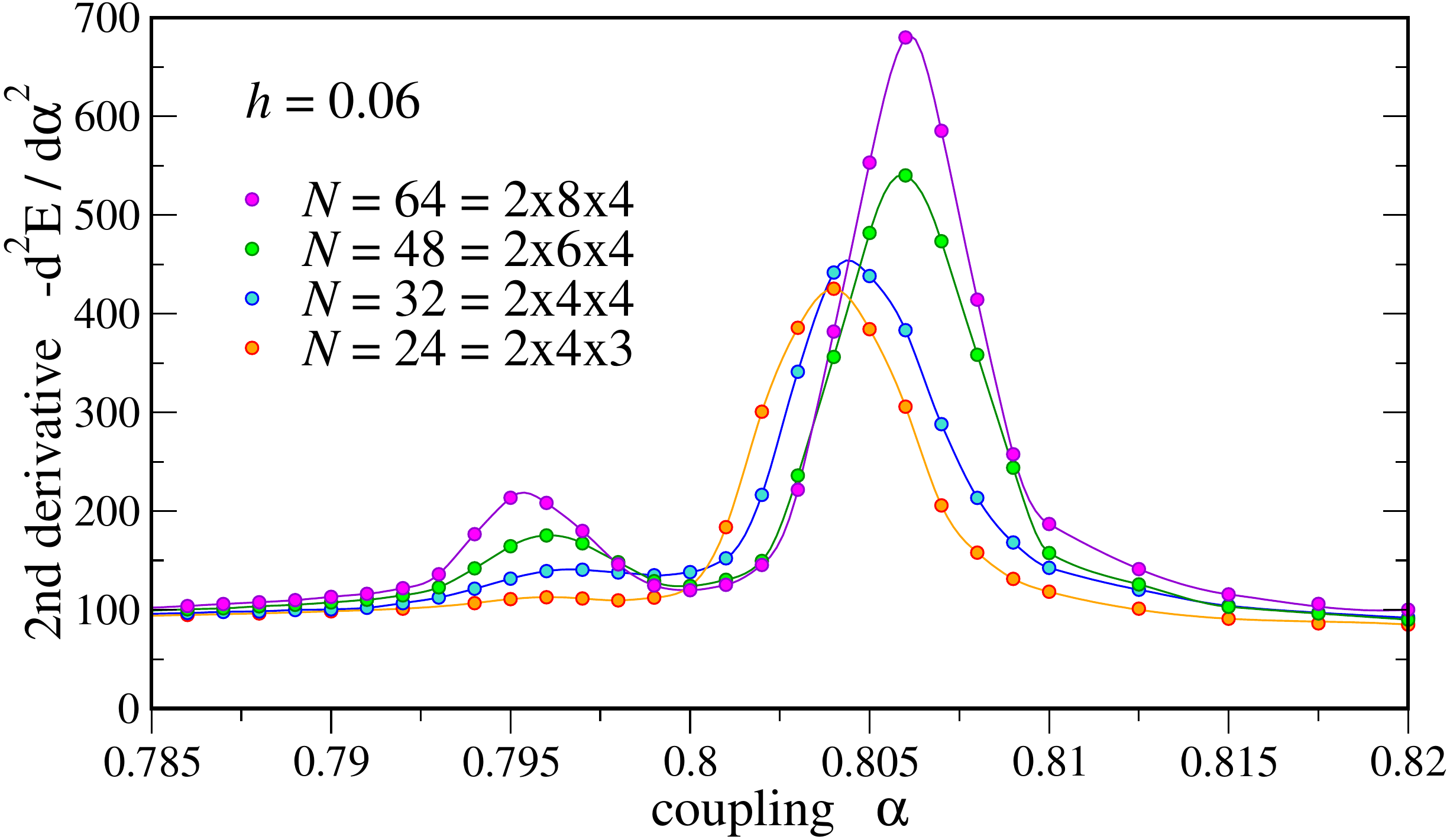}
    }
  \caption{(color online)
              Constant field scan in the vicinity of the (multicritical) point
              separating the stripy AFM from the topological phase.}
  \label{Fig:ScanMultiPoint}
\end{figure}

Finally, we note that there appears to be a second multicritical point
in our phase diagram around $\alpha \approx 0.8$ and $h=0$, where
the stripy AFM phase and the gapless spin liquid meet.
We find that in the presence of the magnetic field the transition lines
of the field-driven phase transition out of the corresponding canted
and topologically ordered states bend in and merge
only in the zero-field limit as depicted in the inset of Fig.~\ref{Fig:Phase_HK}.
To show that there is indeed no direct transition between the canted
stripy N\'eel state and the topological phase we have made extensive
scans in the coupling parameter $\alpha$ in the vicinity of this putative
multicritical point for small field strength.
As shown in Fig.~\ref{Fig:ScanMultiPoint} for $h=0.06$, the second
derivative $d^2E/d\alpha^2$ of the ground-state energy clearly
shows two peaks proliferating with increasing system size indicative
of two well separated phase transitions.
However, the underlying effective theory for such a multicritical point is
not known, and will be left for further study.

\paragraph{Outlook.--}
Having established the rich phase diagram of the Heisenberg-Kitaev model in a magnetic field,
it is interesting to speculate where one would place the Iridate Na$_2$IrO$_3$.
While experiments \cite{Gegenwart} report indications of an AFM ordered ground state below $T_{\rm N} \approx 15$~K,
the precise nature of the order remains open. Given the considerable suppression of the ordering temperature $T_{\rm N}$
in comparison with the Curie-Weiss temperature $\Theta_{\rm CW} \approx 116$~K \cite{Gegenwart},
which is typically interpreted as an indicator of frustration,
an alternative explanation would be the proximity to a quantum critical point,
such as the multicritical point $\alpha \approx 0.8$ in the context of our phase diagram.
This would bring the material in close proximity to the spin liquid phase for $\alpha \gtrsim 0.8$
and the topological phase found for a magnetic field pointing in the \field direction.
To further substantiate this possibility, it is desirable to study the finite-temperature phase diagram
of our model system and to consider the effects of disorder, such as site mixing between
the Ir and Na sites \cite{Gegenwart}.
Finally, it would be interesting to bring the Mott physics discussed in this manuscript in competition
with the topological insulator phase suggested in Ref.~\onlinecite{Shitade09}.

\begin{acknowledgements}
We acknowledge discussions with G. Jackeli, R. Kaul, D. N. Sheng, and R. Thomale. XLQ is supported partly  by the Alfred P. Sloan Foundation.
\end{acknowledgements}



\begin{thebibliography}{99}
\bibitem{FootNoteFieldStrength}
Although the field-driven phase transition might share many similarities for topological states
with Abelian and non-Abelian topological order, there also exist important differences.
For the Abelian $Z_2$ gauge theory, it has been shown that the critical field strength is of
the same magnitude as the single vortex gap and as a result, the closing of the gap for a single
vortex  will naturally lead to vortex condensation.
In the non-Abelian case, however, the single vortex gap, estimated around $\sim0.27$,
is considerably larger than the critical field strength. There are two potential reasons
for this discrepancy in the non-Abelian case:
First, the non-Abelian nature of the vortices gives rise to a macroscopic degeneracy
which might in turn enhance the low energy quantum fluctuations.
Second, in the limit of high vortex density, the vortex core energy can be much smaller
than in the single vortex case and thus further enhance the quantum fluctuations.
\end{thebibliography}
\end{document}